\documentclass[pra,twocolumn,showpacs,aps]{revtex4}

\usepackage{graphicx,color,textcomp}
\usepackage{amsmath}

\usepackage{dcolumn,amsmath,amsthm,amscd,amsfonts,amssymb,epsfig,graphics,graphicx,eucal}
\usepackage{color}

\newcommand{\beq}{\begin{eqnarray}}
\newcommand{\eeq}{\end{eqnarray}}
\newcommand{\beqq}{\begin{eqnarray*}}
\newcommand{\eeqq}{\end{eqnarray*}}

\begin{document}

\title{Stable multidimensional soliton stripes in two-component Bose-Einstein condensates}

\author{Valeriy A. Brazhnyi}

\affiliation{Centro de F\'{\i}sica do Porto, Faculdade de Ci\^encias, Universidade do Porto, R. Campo Alegre 687, Porto 4169-007, Portugal}

\author{V\'{\i}ctor M. P\'erez-Garc\'{\i}a}
\affiliation{Departamento de
Matem\'aticas, E. T. S. de Ingenieros Industriales, and Instituto de Matem\'atica Aplicada a la Ciencia y la Ingenier\'{\i}a, Universidad de Castilla-La
Mancha 13071 Ciudad Real, Spain}


\date{\today}

\begin{abstract}
We discuss how to construct stable multidimensional extensions of one-dimensional dark solitons,  the so-called soliton stripes,  in two-species Bose-Einstein condensates in the immiscible regime. We show how using a second component to fill the core a dark soliton stripe leads to reduced instabilities while propagating in homogeneous media. We also discuss  how in the presence of a trap arbitrarily long-lived dark soliton stripes can be constructed by increasing the filling of the dark stripe core. Numerical evidences of the robustness of the dark soliton stripes in collision scenarios are also provided.
\end{abstract}

\maketitle

\section{Introduction}

 One of the most remarkable achievements in quantum physics in the last decade was the Bose-Einstein condensation of  ultracold alkaline atoms \cite{reviews}. Because of the interatomic interactions present in Bose-Einstein condensates (BECs), they have a high potential for supporting quantum nonlinear coherent excitations of which many types  have been already been experimentally observed: dark  \cite{dark1,dark2}, bright  \cite{bright}  and gap solitons \cite{gap}, vortices \cite{vortices}, vortex rings \cite{vrings},  shock waves \cite{shockwaves}, vector solitons \cite{vector1}, and others \cite{Faraday}. Many other types of nonlinear excitations have been theoretically predicted to exist (see e.g. Refs. \cite{PGR,PK} and references therein).

Although solitons have attracted a lot of attention because of their robustness, they do not survive in effective dimensions higher than one:  bright solitons are unstable to blow-up \cite{blowup,Sulem} and dark soliton stripes  decay into point vortices due to the so-called snake instability \cite{vrings}. The development of this instability in the context of BECs has received a lot of attention \cite{Brand}. 
Other multidimensional extensions of the dark soliton, such as the ring dark soliton \cite{KDS}, are also unstable \cite{Ring_DS}.

In multicomponent Bose-Einstein condensates the extra degrees of freedom provided by the existence of another component allows for the existence of many different types of solitons. In the context of BEC there has been a strong interest on direct extensions of the one-dimensional (1D) solitons, such as dark-dark, or  bright-bright solitons \cite{VD1,VD2,VD3,VD4,VD5} and also on many other types of nonlinear waves that do not have direct 1D analogues such as dark-bright solitons \cite{BA,DB1,DB2}, soliton molecules \cite{Smol}, domain walls \cite{DW1,DW2,DW3}, symbiotic solitons \cite{SS1,SS2}, solitons on plane wave solutions  \cite{SPW}, and supersolitons \cite{SSol1}. However, as it happens in the scalar case most of these solutions are unstable in multidimensional scenarios.

In this paper we will study how adding a second component allows for the construction of stable multidimensional extensions of the dark soliton stripe thus leading to stable dark-bright soliton stripes (DBSS). 
The idea, as in the dark-bright soliton \cite{BA,DB1,DB2} concept, is based on using a second immiscible condensate to fill the dark stripe thus increasing the rigidity of the coupled system. Although different types of nonlinearly coupled structures have been studied in multicomponent BECs (see e.g. Refs. \cite{multi1,multi2} and references therein), to our knowledge this possibility has not been studied in the context of multicomponent trapped BECs.

The structure of the paper is as follows. First, in Sec. \ref{Sec2} we present our model problem and the basic facts in one-dimensional scenarios. In Sec. \ref{Sec3} we discuss the behavior of dark-bright soliton stripes in homogeneous media and discuss their linear stability. In Sec. \ref{Sec4} we discuss the novel features due to the presence of transverse trapping and show how the combination of the stripe filling due to a second component plus the transverse trapping leads to extremely robust dark-bright soliton stripes. This robustness manifests also in dark-bright soliton stripe collisions, studied in Sec. \ref{Sec5}. Finally, in Sec. \ref{Sec6} we summarize our conclusions.

\section{Model equations and dark-bright solitons}
\label{Sec2}

\subsection{Model equations  and parameters}

We will consider a quasi-two dimensinal dilute binary mixture of BECs in a pancake 
trap described in the mean field limit by the adimensional equations
\begin{equation}
i\frac{\partial\psi_j}{\partial t} = \left[-\frac{1}{2} \triangle  + V(x,y) \right]\psi_j  +\left(\sum_{k} g_{jk}|\psi_k|^2 \right) \psi_j,
\label{NLS_GP}
\end{equation}
where $j=1,2$ and $\triangle = \partial^2/\partial x^2 + \partial^2/\partial y^2$. The wavefunction $\psi_j$ is related to the one in physical units (denoted by tilde) in the full space through 
\begin{equation}
\Psi_j (\tilde x, \tilde y, \tilde z, \tilde t)=\frac{1}{\left(2\sqrt{2\pi} l_z^2 a_{11}\right)^{1/2}}\psi_j (x,y) \varphi_0(z)e^{-i  t/2},
\end{equation}
 where $\Psi_j$ are normalized to the numbers of atoms $N_j$: $\int|\Psi_j|^2d{\bf r}= N_j$, and $\varphi_0(z)=(1/\pi)^{1/4}\exp(- z^2/2)$, ${\bf r}=\tilde {\bf r}/l_z$, $ t=\tilde t\omega_z$ where $l_z=\sqrt{\hbar/m\omega_z}$ and $\omega_z$ is the trap frequency along $z$ direction corresponding to a trapping potential $V(\tilde z) = m\omega_z^2\tilde z^2/2$ in the original model written in physical units. 
Finally, the nonlinear coefficients are defined through $g_{jk}=a_{jk} /a_{11}$, where $a_{11}$ is the intra-component scattering length of the first component (assumed, as all of the others, to be positive).

For simplicity, all the examples to be discussed in this paper will refer to the case of equal inter- and intra-species interactions: $g_{11}=g_{12}=g_{21}=g_{22}= 1$, that is not far from several physically relevant situations. We have checked that our ideas to be presented in this paper remain qualitatively valid for a wide range of parameter combinations corresponding to two-component BECs in the immiscible regime.

\subsection{Dark-bright solitons}

Let us first consider the situation $V(x,y) = 0$ and take initial data of the form 
\begin{subequations}
\label{id}
\begin{eqnarray}
\psi_{1}(x,y,t) & = & \psi_{d}(x,t)f(y), \\
\psi_{2}(x,y,t) & = & \psi_{b}(x,t)f(y),
\end{eqnarray}
\end{subequations}
where  $\psi_d$, $\psi_b$ solve the coupled NLS equations~\cite{BA}
\begin{subequations}
\label{GP}
\begin{eqnarray}
i\frac{\partial\psi_d}{\partial t} &= &-\frac12\frac{\partial^2 \psi_d}{\partial x^2} + \left(\left|\psi_d\right|^2 + 
\left|\psi_b\right|^2 - \rho \right) \psi_d, \label{GP1} \\
i\frac{\partial\psi_b}{\partial t} &= &-\frac12\frac{\partial^2 \psi_b}{\partial x^2} + \left(\left|\psi_b\right|^2 + 
\left|\psi_d\right|^2 - \rho-\delta \right) \psi_b. \label{GP2}
\end{eqnarray}
\end{subequations}
Eqs. (\ref{GP})  have exact  dark-bright soliton solutions \cite{BA} \begin{subequations}\label{DBs}
\begin{eqnarray}
\psi_d(x,t) &=& \sqrt{\rho}\left[i \sin\alpha +\cos\alpha \tanh\beta\left(x-X(t)\right)\right],
\label{dark_sol}\\
\psi_b(x,t) &=& A_b {\rm sech}\beta(x-X(t)) e^{ i(\phi+x\beta \tan\alpha +\Omega_b t)},
\label{bright_sol}
\end{eqnarray}
\end{subequations}
where $A_b = \sqrt{N_b\beta/2}$, $\beta=\sqrt{\rho \cos^2\alpha + (N_b/4)^2}-N_b/4$ with $N_b=\int |\psi_b|^2 dx$ and $\Omega_b=\beta^2(1-\tan^2\alpha)/2-\delta$ (in what follows we will take $\delta=0$). 
The soliton position is determined by the equation $X(t)=x_0+t\beta\tan\alpha$.
In the scalar case the parameter $\alpha$ is related to the dark soliton velocity as $v=\sqrt\rho\sin \alpha$, but in the vector case  due to the presence of the bright component the velocity of the vector soliton for the same $\alpha$ is slower, and given by the formula  $v_{db}=\beta\tan\alpha$.

The bright component also has its own global phase $\phi$ that is relevant only when interactions between two bright solitons are considered.

It is worth to mention that this dark-bright soliton exists only in the immiscible regime of repulsively interacting systems and has nothing to do with the bright solitons existing in systems with attractive interactions. Through this paper when speaking of ``bright solitons" or ``the bright component" we will refer to 
the repulsively interacting component that is hosted by the dark soliton.

\section{Dark-bright soliton stripes and their stability} 
\label{Sec3}

Taking $f(y) = 1$ in Eq. (\ref{id}) an exact solution of Eqs. (\ref{NLS_GP}) with appropriate boundary conditions in two spatial dimensions is given by Eq. (\ref{DBs}). 

In the scalar case, this dark soliton stripe is known to be unstable \cite{Turitsyn}. 
Here we will check stability of the DBSS considering Eqs. (\ref{NLS_GP}) and taking the bright component as unperturbed, namely  $\psi_b(x,y)=\psi_{b0}(x)$  while the dark component is subject to the transverse perturbation  
\begin{equation}
\psi_d(x,y)=\psi_{d0}(x)+a(x)e^{-iky+i\Omega t}+b^*(x)e^{iky-i\Omega t}.
\end{equation}

\begin{figure}[ht]
\epsfig{file=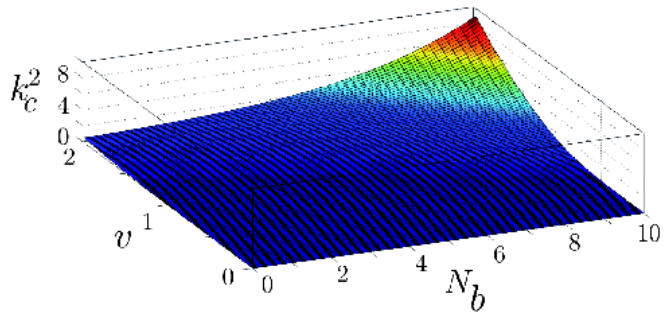,width=\columnwidth}
 \caption{Surface plot of the critical wavenumber $k_c$  for which the instability appears as a function of the relevant parameters $(w,N_b)$.} \label{stab}
\end{figure}


Substituting $\psi_{b,d}$ into Eqs. (\ref{NLS_GP}) in the frame moving with the velocity of the vector soliton, $v_{db}$, and decomposing $a(x)$ and $b(x)$ into real and imaginary parts $a=f_1+if_2$ and $b=f_1-if_2$ we get
\begin{equation}
\left(i\Omega \sigma - \frac12 k^2\right) F + L F=0, 
\end{equation}
where $F= \begin{pmatrix} f_1\\ f_2\end{pmatrix}$, $\sigma=\begin{pmatrix} 0& 1 \\ -1 & 0 \end{pmatrix}$, and 
\begin{equation}
L =L_1  -  \begin{pmatrix} P  -Q^+ & i Q^- \\  i Q^- &  P+ Q^+ \end{pmatrix},
\label{syst_f}
\end{equation}
being 
\begin{subequations}
\begin{eqnarray}
L_1 & = &  v_{db}\sigma\partial_x + \tfrac12\partial_{xx}, \\ P & = & 2\left|\psi_{d0} \right|^2+\left|\psi_{b0} \right|^2 - \rho, \\ Q^{\pm} & =&-  \tfrac 12\left(\psi_{d0}^2\pm (\psi_{d0}^*)^2 \right).
\end{eqnarray}
\end{subequations}
Rewriting the unperturbed solution given by Eq. \eqref{DBs} in the moving frame, defining 
 $w=\sqrt\rho \cos \alpha$, rewriting $\beta$ and $v_{db}$ in terms of $w$ and inserting  $\psi_{d0}$ and $\psi_{b0}$ in Eqs. (\ref{syst_f}) we find
\begin{equation}
L= L_1 - \begin{pmatrix} \frac{6w^2-N_b\beta}{2\cosh^2(\beta x)} - 2w^2 & -2vw\tanh(\beta x) \\ -2vw\tanh(\beta x)   & \frac{2w^2-N_b\beta}{2\cosh^2(\beta x)} - 2v^2 \end{pmatrix}.
\label{L}
\end{equation}
The eigenvalue problem can be rewritten in the form 
\begin{equation}
L F=-\tfrac 12 p^2 F,
\end{equation}
where $\tfrac12 p^2=i\Omega\sigma-\frac 12 k^2$ and $p^2$ plays the role of a new eigenvalue. Finally, after introducing the rescaling 
\begin{equation}
\left\{\Omega/\beta^2,  k/\beta, \beta x, p/\beta, v/\beta,  w/\beta \right\}=\left\{\tilde \Omega, \tilde k, \tilde x, \tilde p,\tilde v,\tilde w\right\}
\end{equation}
and looking for rescaled eigenfuctions  of the form
\begin{equation}
f_1=e^{iq\tilde x} g_1(\tilde x), \qquad f_2=e^{iq\tilde x} g_2(\tilde x),
\end{equation}
 where $g_1(\tilde x)$ and $g_2(\tilde x)$ are unknown functions which are constant in the limit $|x|\to \infty$, 
 of the rescaled eigenvalue problem 
\begin{equation}
\tilde L F=-\frac 12 \tilde p^2 F,
\end{equation} we obtain, after some algebra
\begin{equation}
\begin{pmatrix} - 2\tilde w^2- \frac12 q^2 +\frac 12 \tilde p^2 & -2\tilde v \tilde w+iq \tilde v/\tilde w \\ -2\tilde v \tilde w -iq \tilde v/\tilde w   &  - 2\tilde v^2 - \frac12 q^2 +\frac 12 \tilde p^2 \end{pmatrix} \begin{pmatrix} g_1\\ g_2\end{pmatrix} =0 .
\label{tL}
\end{equation}
\begin{figure}[ht]
\epsfig{file=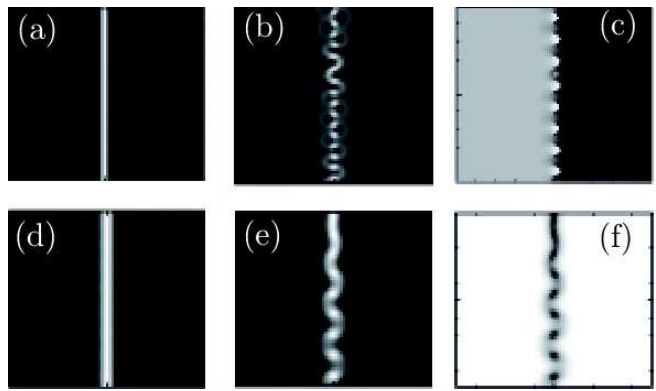,width=\columnwidth}
 \caption{Density plots of the dynamics of a perturbed dark soliton stripe with a homogeneous background. (a-c) Scalar case, i.e. $N_b=0$. Shown are the densities $|\psi(x,y,t)|^2$ for (a) $t=0$ and (b) $t=180$ and (c) the phase $\text{arg}(\psi)$ for $t=180$. (d-f) Vector case with a bright component $N_b=3$. Shown are the densities $|\psi_d(x,y,t)|^2$ for (d) $t=0$ and (e) $t=570$ and (f) the second component $|\psi_b(x,y,t)|^2$ for $t=570$. Parameters are $\rho=1$, $\lambda=0$, $\kappa=0$ and the spatial region shown in the all the subplots is $(x,y) \in [-40,40]\times [-40,40]$. The initial stationary profile was perturbed by taking $\psi_j=\psi_j(t=0)(1+0.005\cos(0.1y))$} 
\label{homo}
\end{figure}

Solving the eigenvalue problem (\ref{tL}) we get the relation between $\tilde p$ and $\tilde q$
\begin{equation}
\tilde p^2 = 2(\tilde v^2+\tilde w^2) + q^2 \pm 2\sqrt{(\tilde v^2+\tilde w^2)^2 + \tilde v/(\tilde w q^2)}.\label{pq}
\end{equation}
It can be shown (see \cite{Turitsyn}) that an eigenvalue from the discrete spectrum of the operator $L$ corresponding to a bound state with $\Omega=0$ is $q=i$ what leads to the following critical wavenumber  below which the long-wavelength instability appears in the non-scaled variables
\begin{eqnarray}
k_c^2 = \beta^2-2 +  2\sqrt{\beta^4(1-1/w^2) + 1}.
\label{kc}
\end{eqnarray}

In Fig. \ref{stab} we plot the dependence of the critical wavenumber $k_c$ on the relevant parameters $(w,N_b)$. It is obvious that the presence of a nonzero bright component, although unable to completely inhibit the instability, reduces significantly the range of unstable wavenumbers.

This analysis is confirmed by direct numerical simulations of Eqs. (\ref{NLS_GP}) for the evolution of a black stripe on a homogeneous background. Some results are summarized in Fig. \ref{homo}. It is clearly seen how introducing a bright component leads to a substantial increase in the time at which the snake instability sets in. For a small bright component with $N_b=3$, the DBSS shown in Fig. \ref{homo}(d) (notice the wider dark stripe in comparison with Fig. \ref{homo}(a) due to the presence of the bright component) becomes unstable for $t\simeq 570$, these times being three times longer than the instability times found for $N_b=0$. By increasing the number of atoms in the bright component it is possible to get stable dynamics for much longer times.

\section{Waveguide configurations} 
\label{Sec4}

\begin{figure}[ht]
\epsfig{file=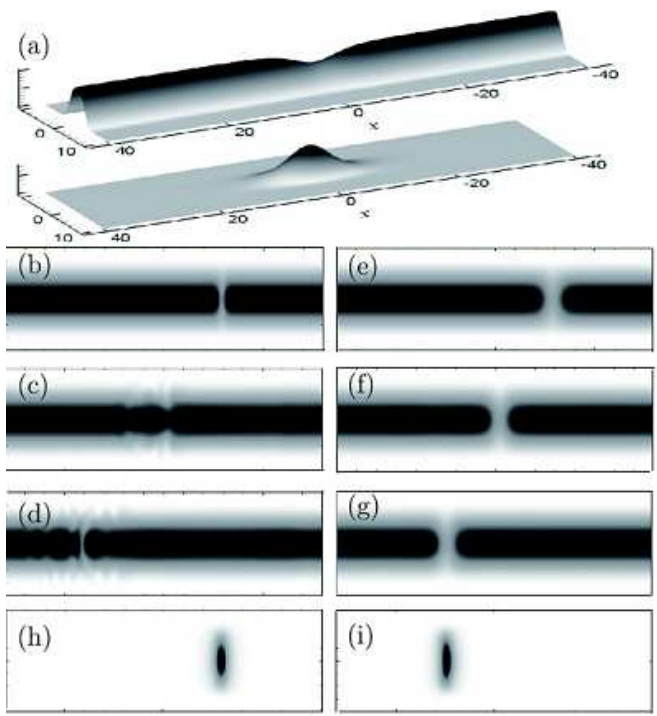,width=\columnwidth}
\caption{Simulation of the evolution of a DBSS. 
(a) Two-dimensional surface plots of a DBSS with initial data, $|\psi_d(x,y,0)|^2$, $|\psi_b(x,y,0)|^2$  given by Eqs. (\ref{id}), with $f$ given by Eq. (\ref{BG_f}) for  parameter values $\rho=1$, $\kappa=0.4$,  $N_b=3$, $\phi=0$, and $v=-0.6$. 
(b-i) DBSS propagation for parameter values  $\rho=1$, $\kappa=0.1$, $v=-0.6$. Shown are density plots of $|\psi_d(x,y,t)|^2$ for (b-d) $N_b = 0$ and times (b) $t=0$, (c) $t=50$, (d) $t=100$  and for $N_b=3$ (e-g) for times (e) $t=0$, (f) $t=100$ and (g) $t=200$. Density plots of the bright component in the multicomponent case are also shown in subplots (h-i) for $t=0$ (h) and $t=200$ (i). The spatial region shown in subplots (b-i) is $x\in [-80,80], y\in [-20,20].$
} \label{fig_iniprof}
\end{figure}

Let us now consider more realistic scenarios by including a transverse trapping potential
in a waveguide geometry taking $V(y)= \kappa^2 y^2/2$. 

Due to the spatial inhomogeneities the analytical stability analysis is intractable and we will base our analysis on numerical simulations. We will take the transverse part of the solution $f(y)$ in Eq. (\ref{id}) as given by the solution of the 
nonlinear spectral problem for $(\mu, f_{\mu})$
\begin{equation}
\frac12\frac{\partial^2 f}{\partial y^2} + \left[ \mu - \frac12\kappa^2y^2\right]f -\rho f^3	=0,
 \label{BG_f}
\end{equation}
with boundary conditions $f(0)=1$, $f_y (0)=0$ and $\lim_{y\to + \infty}f(y)=0$.
A typical initial profile for the waveguide configuration calculated from  Eqs. (\ref{DBs}) and (\ref{BG_f}) is shown in Fig. \ref{fig_iniprof}(a).

\begin{figure}[ht]
\epsfig{file=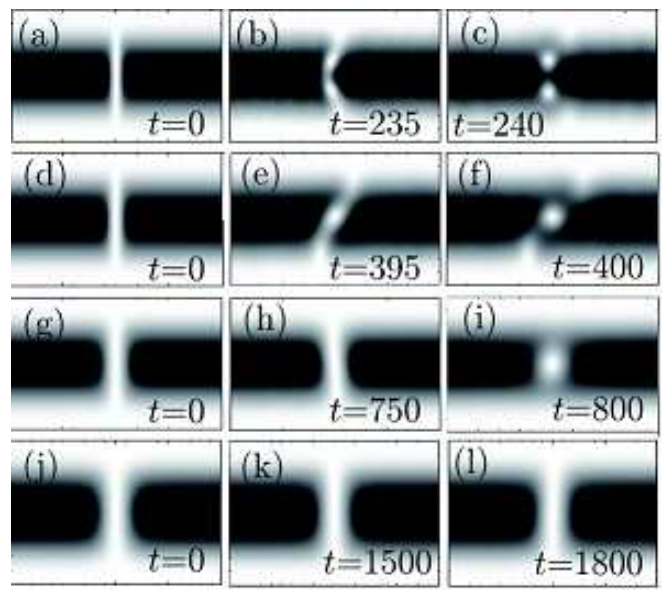,width=\columnwidth}
 \caption{Density plots of the results of the simulation of the transition to more stable behavior of DBSSs as the number of particles in the bright component $N_b$ increases for parameter values $\kappa = 0.2$, (a-c) $N_b = 0$, (d-f) $N_b=1$, (g-i) $N_b = 3$, (j-l) $N_b = 5$. The times are indicated on each subplot. The spatial regions shown in all of the subplots is $(x,y) \in [-20,20] \times [-10,10]$.
} \label{fig_transition}
\end{figure}

We have used these solutions as initial data for a numerical study of the stability of trapped DBSSs. In all of our simulations to be presented in this section, the condensates are far from the 1D limit since in that limit dark solitons are stable and our dark stripes are all unstable when their cores are empty.
In this case, and due to the previously discussed enhancement of the stability due to the filling of the dark soliton stripe it is possible to make an unstable scalar configuration to be stable, leading to \emph{stable multidimensional DBSSs}. Fig. \ref{fig_iniprof}(e-g) shows an example of this behavior. 

In Fig. \ref{fig_iniprof}(b-d) we can see that the scalar dark soliton stripe becomes curved when propagating in the waveguide, a well known effect due to the dependence of the soliton velocity on the (now inhomogeneous) background amplitude. Thus in the low amplitude regions the stripe moves more slowly and transforms into smaller amplitude gray solitons accompanied by the formation of an array of vortices on the tails of the radial density distribution. This behavior is confirmed by the analysis of the phase distribution of the dark component.

It is known that the transverse trapping plays a stabilizing role in the dynamics of multidimensional solitons that in a certain tight-binding regime become effectively 1D and thus stable \cite{Brand}.
This fact is also in agreement with the ideas of Ref. \cite{BA} where it was suggested that in higher dimensions dark-bright solitons should be more stable than pure dark solitons, even in traps that do not attain the quasi-1D regime. 
However our results go beyond that since we have tested many different configurations including very weakly trapping potentials and have found that it is allways possible to find a filling that makes the DBSS stable for very large times. The transition from the unstable regime to the stable one as the number of particles in the bright component is increased is very clear, the number of vortices generated being smaller and the instability setting in later as a function of $N_b$. 

An example of that behavior is shown in Fig. \ref{fig_transition}. It can be seen how for a given trapping, the scalar dark soliton stripe becomes unstable and breaks into four point vortices [Fig. \ref{fig_transition}(a-c)], as the bright component is increased to $N_b=1$ [Fig. \ref{fig_transition}(d-f)], 
$N_b=3$ [Fig. \ref{fig_transition}(g-i)], and $N_b=5$ [Fig. \ref{fig_transition}(j-l)],  the number of vortices arising after the instability develops are reduced 
to three, one and none, respectively.  The times required for the instability to set in become larger with increasing $N_b$ as discussed previously. For $N_b=5$ the DBSS is stable for the very long times used in our simulations.

\section{Interaction between dark-bright soliton stripes}
\label{Sec5}

Robustness in soliton-solitons interaction is one of the key elements of these nonlinear waves. In our case, despite our 2D dark-bright stripe structures studied are not strict (integrable) solitons, they display collisional properties that support their robustness. In this section we will consider fast and slow DBSSs collisions as well as their interactions in ring trap configurations.

\subsection{Fast collisions}

\begin{figure}[ht]
\epsfig{file=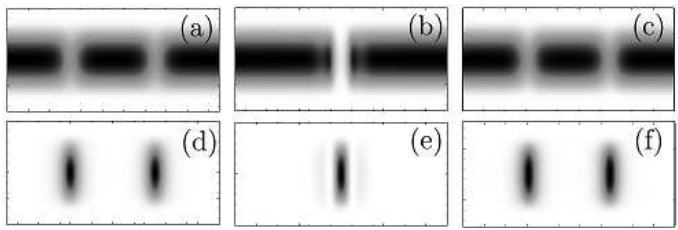,width=9cm}
\caption{Interaction of two DBSSs separated initially in space by a distance $\xi=40$. 
Parameters are the same for both solitons: $\rho=1$, $|\alpha|=0.6$,  $\kappa=0.1$, $\Delta\phi=0$, $\xi=20$, $N_b = 3$. 
The upper/bottom row corresponds to the dark/bright components  for  (a) $t=0$, (b) $t= 70$, and (c) $t= 140$, respectively. The spatial region shown is $x\in [-50,50], y\in [-20,20].$
} \label{fig_2DB_06}
\end{figure}

First, in fast head-on collisions DBSSs show clearly their robustness. We have simulated many collision scenarios and the typical outcome is a pair of outgoing DBSSs with the usual time delay due to the intreraction. A typical simulation is shown in Fig. \ref{fig_2DB_06}.

\subsection{Ring trap configurations}

As another demonstration of the robustness of the DBSSs, we have simulated numerically their propagation and collisions in toroidal traps \cite{exrt1,exrt2}. These configurations have been suggested to provide interesting test beds for the propagation of different types of solitons \cite{rt1,rt1a,rt1b,rt2}. In our calculation we adapt the waveguide configuration of the trap by using a trapping potential of the form
 \begin{equation}
 V(x,y)=\nu^2 \left(r-R\right)^2/2,
 \end{equation}
  where $R$ is ring radius and ${\bf r}=\sqrt{x^2+y^2}$.
A typical  example is shown in Fig. \ref{fig_DB_tor} were both the stable propagation of two dark-bright solitons along the toroidal BEC and their stable behavior after the collision, returning to the initial state for $t=1000$ are shown. 

\begin{widetext} \hbox{}
\begin{figure}[ht]
\epsfig{file=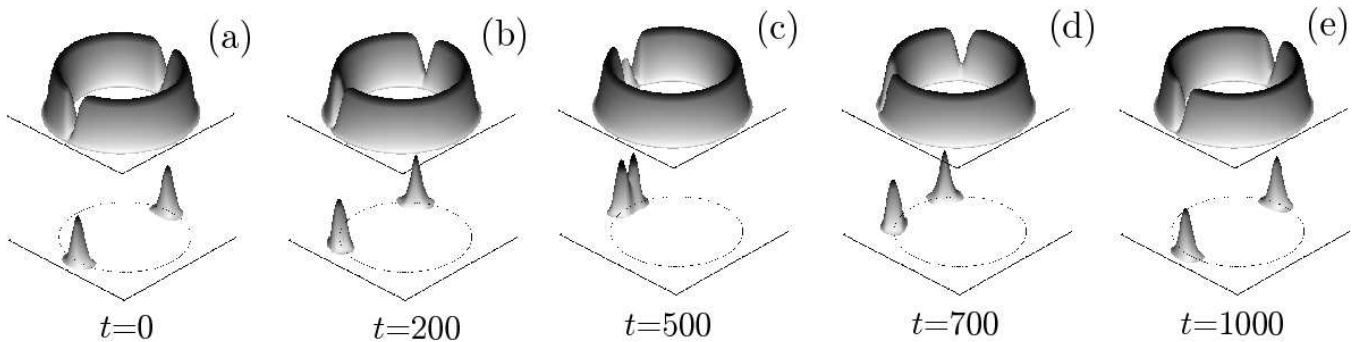,width=\textwidth}
 \caption{Collision of DBSSs in a toroidal trap. The soliton and trap parameters are $\kappa=0.2$, $R=40$, $N_b=6$, $v=0.9$. Shown are surface plots of the dark (upper row) and bright components (lower row) on the spatial region $(x,y) \in [-50,50] \times [-50,50]$ for times: (a) $t=00$, (b) $t=200$, (c) $t=500$, (d) $t=700$ and (e) $t=1000$.} \label{fig_DB_tor}
\end{figure}
\end{widetext}

\subsection{Slow collisions}

As it can be expected \emph{a priori}, slow (low energy)  DBSSs collisions lead to much more complex dynamics. 
To show it in this subsection  we 
will consider the interaction of two DBSSs with zero initial velocity (black dark components) initially separated by a distance $\xi$. 
In this case the phase difference between the bright components, i.e. $\Delta\phi=\phi_1-\phi_2$ plays an important role. 
When  the phase difference is zero ($\Delta \phi=0$) the bright components in the DBSSs repel each other while when this difference is $\Delta \phi=\pi$ they have an attractive interaction as it happens in the scalar case. As to the dark components, their interaction is always repulsive.

\begin{figure}[ht]
\epsfig{file=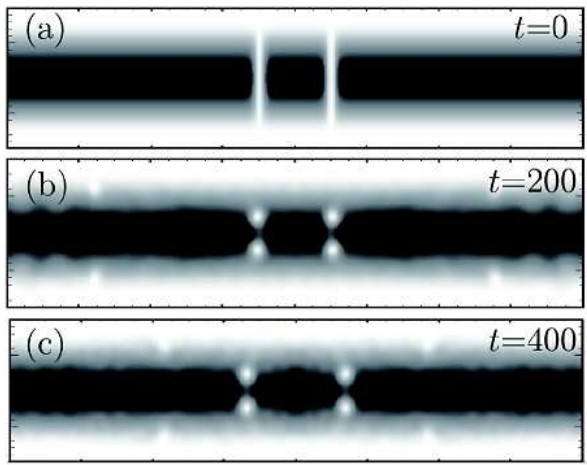,width=0.9\columnwidth}
\caption{Interaction of two DBSSs separated initially in space by a distance $\xi=20$. 
Parameters are the same for both solitons: $\rho=1$, $\alpha=0$,  $\kappa=0.1$, $\Delta\phi=0$, $\xi=20$.
$N_b = 1$. 
The spatial region shown in all of the subplots is $x\in [-80,80], y\in [-20,20].$
} \label{fig_2DB_inter-a}
\end{figure}

\begin{figure}[ht]
\epsfig{file=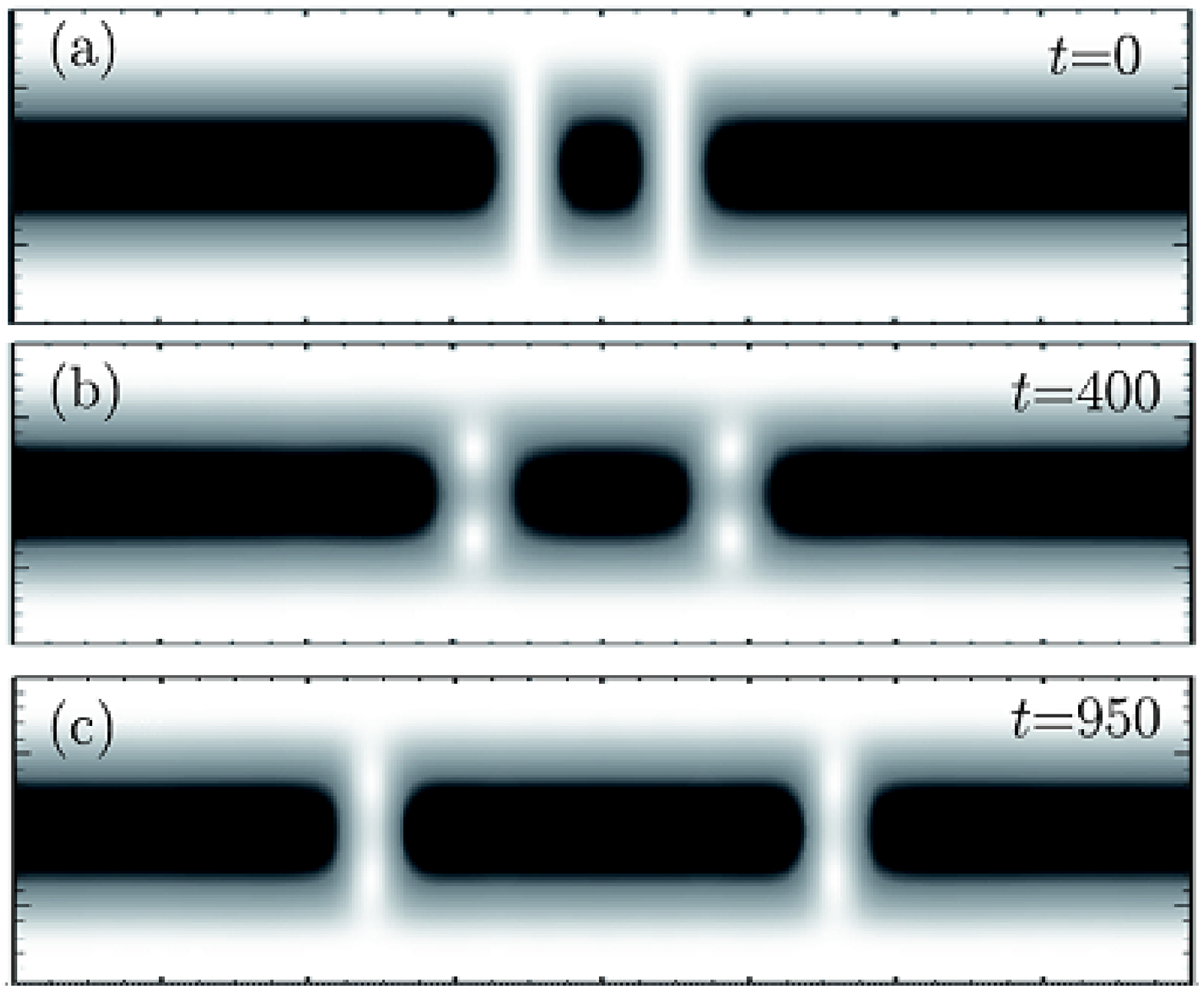,width=0.9\columnwidth}
\caption{
Same as in Fig. \ref{fig_2DB_inter-a} but for $N_b=5$.
} 
\label{fig_2DB_inter-b}
\end{figure}

\begin{figure}[ht]
\epsfig{file=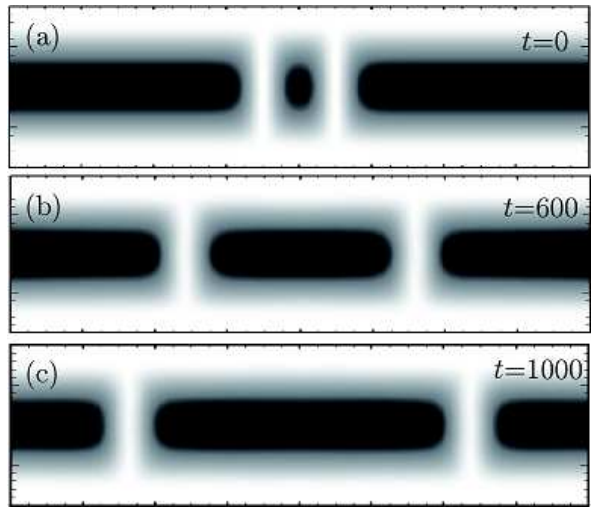,width=0.9\columnwidth}
\caption{Same as in Fig. \ref{fig_2DB_inter-a} but for $N_b=8$.
} \label{fig_2DB_inter-c}
\end{figure}

We present an example of our numerical experiments in  Fig. \ref{fig_2DB_inter-a}. 
For small bright components, $N_b=1$ (see Fig. \ref{fig_2DB_inter-a}) and due to the repulsive force between the dark soliton stripes two pairs of vortices are excited and counter-propagate (see the outer part of the figure) while another two pairs also repel each other but move more slowly. Increasing the bright component to $N_b=5$ (Fig. \ref{fig_2DB_inter-b}) leads to a more stable propagation although still some profile oscillations and vortex generation are observed. Finally, using a larger bright component $N_b=8$ leads to a perfectly stable propagation of the DBSSs (Fig. \ref{fig_2DB_inter-c}).

\begin{figure}[ht]
\epsfig{file=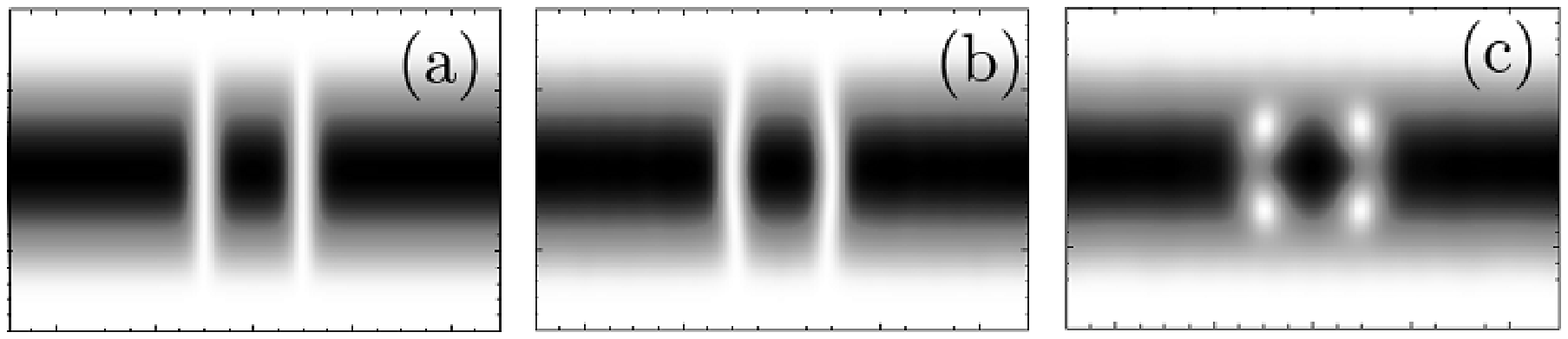,width=9cm}
\epsfig{file=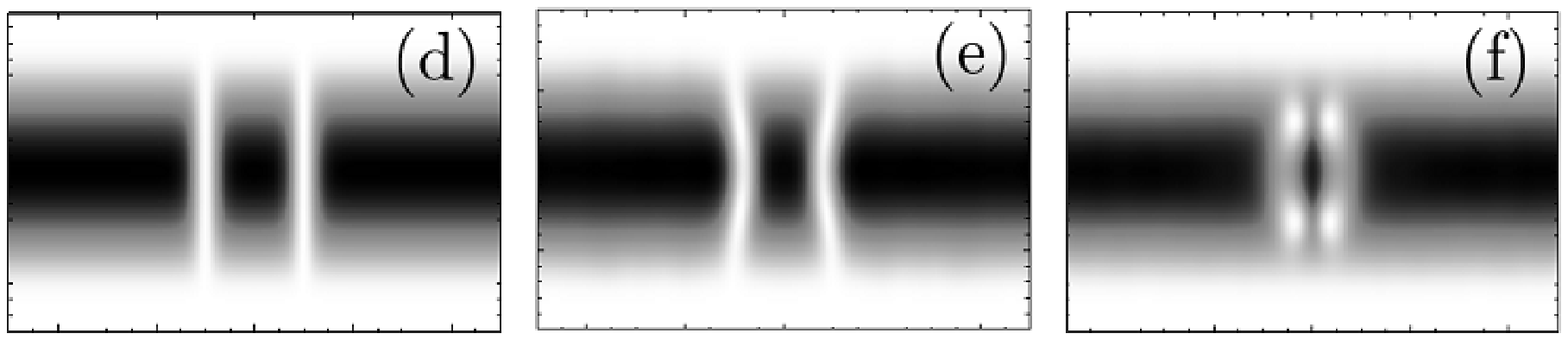,width=9cm}
\caption{Interaction of two DBSSs separated in space by a distance $\xi=20$. 
Parameters are the same for both solitons: $\rho=1$, $\alpha=0$,  $\kappa=0.1$, $\Delta\phi=\pi$, $\xi=20$.
(a-c) $N_b = 3.6$ ($t=0; 150; 200$), (d-f) $N_b=3.7$ ($t=0; 220; 360$). The spatial region shown in all of the subplots is $x\in [-50,50], y\in [-20,20].$
} \label{fig_2DB_pi}
\end{figure}

\begin{figure}[ht]
\epsfig{file=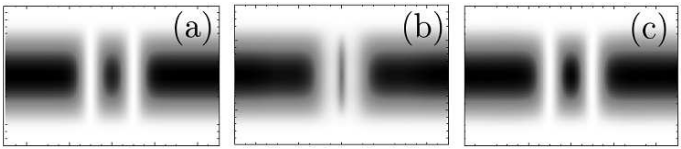,width=9cm}
\caption{Same as in Fig. \ref{fig_2DB_pi} but for $N_b=8$. Shown are density plots of the dark component for times  (a) $t=0$, (b) $t=430$ and (c) $t= 600$.} \label{fig_2DB_Nb8_pi}
\end{figure}

By changing the interaction between the bright components from repulsive ($\Delta \phi=0$) to attractive  ($\Delta\phi=\pi$) it can overcome the repulsive interaction between the dark components leading to a direct collision between the DBSSs. The outcome of such collisions is shown in Fig. \ref{fig_2DB_pi}. 
We will study two situations, the first summarized in Fig. \ref{fig_2DB_pi}(a-c) where the bright component is below a critical value of $N_b$ and the attractive interactions are not sufficient to overcome the repulsive force coming from the interaction between the dark stripes. The second case studied is summarized in Fig. \ref{fig_2DB_pi}(d-f), where the attraction leads to head-on collision between both DBSSs. 
In both cases the interaction leads to the destabilization of the dark soliton stripes. This is due to the fact that in these types of slow collisions marginally stable multidimensional solitons have a lot of time to interact and thus the outcome in many cases is the destabilization of the structure \cite{Montesinos,Smol}.  
To overcome instability effects we have increased the bright component and observed periodic collisions between two DBSSs. In Fig. \ref{fig_2DB_Nb8_pi} we show the first period of the oscillations observed in that scenario. However, in this very unstable situation even with this strong bright component after the third collision (period) the instability sets in and DBSSs break into two pairs of vortices propagating outwards (a similar phenomenon with different types of solitons was discussed in Ref. \cite{Montesinos}). 

\section{Conclusions}
\label{Sec6}

In this paper we have discussed the stabilization effect of a bright component on dark soliton stripes in multicomponent BECs in the immiscible regime. 

We have shown using analytical tools and numerical simulation that dark-bright soliton stripe complexes are less unstable in homogeneous systems than their dark soliton stripes counterparts without filling. We have also studied the combined effect of trapping and filling with a second component, that leads  to stable dark-bright soliton stripes. Finally we have presented some examples of the robustness of these dark-bright solitonic stripes in collision scenarios.

\section{Acknowledgements}

The work of V.M. P.-G. has been partially supported by grants
FIS2006-04190  (Ministerio de 
Ciencia e Innovaci\'on, Spain)  and PCI-08-0093
(Junta de Comunidades de Castilla-La Mancha, Spain). V.A.B. acknowledges the support from the FCT grant,
PTDC/FIS/64647/2006.


\end{document}